\documentclass[twocolumn]{aastex63}
\usepackage{courier}
\usepackage[T1]{fontenc}
\usepackage{gensymb}
\usepackage{amssymb}
\usepackage{amsmath}
\usepackage{url}
\DeclareUnicodeCharacter{2212}{-}

\usepackage{mathtools}
\usepackage{changepage} 
\usepackage{url}
\usepackage{appendix}
\usepackage{lineno}

\title{NuSTAR-XMM}

\newcommand{\cha}{\textit{Chandra} }
\newcommand{\bat}{\textit{Swift}-BAT }
\newcommand{\nus}{\textit{NuSTAR} }
\newcommand{\xmm}{\textit{XMM-Newton} }
\newcommand{\bor}{\texttt{borus02} }
\newcommand{\myt}{\texttt{MYTorus }}
\newcommand{\uxc}{\texttt{UXClumpy} }

\begin{document}

\title{Compton-thick AGN in the \textit{NuSTAR} Era. IX. A Joint \textit{NuSTAR} and \textit{XMM-Newton} Analysis of Four Local AGN}

\author{R. Silver}
\affiliation{Department of Physics and Astronomy, Clemson University,  Kinard Lab of Physics, Clemson, SC 29634, USA}

\author{N. Torres-Alb\`{a}}
\affiliation{Department of Physics and Astronomy, Clemson University,  Kinard Lab of Physics, Clemson, SC 29634, USA}

\author{X. Zhao}
\affiliation{Center for Astrophysics $|$ Harvard \& Smithsonian, 60 Garden Street, Cambridge, MA 02138, USA}
\affiliation{Department of Physics and Astronomy, Clemson University,  Kinard Lab of Physics, Clemson, SC 29634, USA}

\author{S. Marchesi}
\affiliation{INAF - Osservatorio di Astrofisica e Scienza dello Spazio di Bologna, Via Piero Gobetti, 93/3, 40129, Bologna, Italy}
\affiliation{Department of Physics and Astronomy, Clemson University,  Kinard Lab of Physics, Clemson, SC 29634, USA}

\author{A. Pizzetti}
\affiliation{Department of Physics and Astronomy, Clemson University,  Kinard Lab of Physics, Clemson, SC 29634, USA}

\author{I. Cox}
\affiliation{Department of Physics and Astronomy, Clemson University,  Kinard Lab of Physics, Clemson, SC 29634, USA}

\author{M. Ajello}
\affiliation{Department of Physics and Astronomy, Clemson University,  Kinard Lab of Physics, Clemson, SC 29634, USA}

\author{G. Cusumano}
\affiliation{INAF-Istituto di Astrofisica Spaziale e Fisica Cosmica, Via Ugo la Malfa, 153, I-90146 Palermo PA, Italy}

\author{V. La Parola}
\affiliation{INAF-Istituto di Astrofisica Spaziale e Fisica Cosmica, Via Ugo la Malfa, 153, I-90146 Palermo PA, Italy}

\author{A. Segreto}
\affiliation{INAF-Istituto di Astrofisica Spaziale e Fisica Cosmica, Via Ugo la Malfa, 153, I-90146 Palermo PA, Italy}

\begin{abstract}
    We present the results of the broadband X-ray spectral analysis of simultaneous \nus and XMM-\textit{Newton} observations of four nearby Compton-thick active galactic nuclei (AGN) candidates selected from the \textit{Swift}-Burst Alert Telescope (BAT) 150-month catalog. This work is part of a larger effort to identify and characterize all Compton-thick (N$_{\rm H}$ $\geq$ 10$^{24}$ cm$^{-2}$) AGN in the local Universe ($z \leq$ 0.05). We used three physically motivated models -- \texttt{MYTorus}, \texttt{borus02}, and \texttt{UXClumpy} -- to fit and characterize these sources. Of the four candidates analyzed, 2MASX J02051994$-$0233055 was found to be an unobscured (N$_{\rm H}$ $<$ 10$^{22}$ cm$^{-2}$) AGN, 2MASX J04075215$-$6116126 and IC 2227 to be Compton-thin (10$^{22}$ cm$^{-2}$ $<$ N$_{\rm H}$ $<$ 10$^{24}$ cm$^{-2}$) AGN, and one, ESO 362$-$8, was confirmed to be a Compton-thick AGN. Additionally, every source was found to have a statistically significant difference between their line-of-sight and average torus hydrogen column density, further supporting the idea that the obscuring material in AGN is inhomogeneous. Furthermore, half of the sources in our sample (2MASX J02051994$-$0233055 and 2MASX J04075215$-$6116126) exhibited significant luminosity variation in the last decade, suggesting that this might be a common feature of AGN.
\end{abstract}

\section{Introduction}
Active Galactic Nuclei (AGN) are supermassive black holes in the center of galaxies that accrete gas from their surrounding material. It is believed AGN are responsible for creating the majority of the cosmic X-ray background (CXB), the diffuse emission observed from 1 to 200$-$300\,keV \citep[e.g., ][]{Alexander_2003, Gilli2007, Treister_2009, Ueda_2014, Brandt2021CXB}. Particularly, a significant fraction \citep[15-20\%;][]{Gilli2007, Ananna2019} at the peak of the CXB \citep[$\sim$30\,keV,][]{Ajello_2008} emanates from a population of AGN with line-of-sight obscuring column densities N$_{\rm H, l.o.s.}$ $\geq$ 10$^{24}$ cm$^{-2}$, known as Compton-thick AGN (CT-AGN). Moreover, population synthesis models, created to properly explain the origins of the CXB, predict CT-AGN comprise between 20\% \citep{Ueda_2014} and 50\% \citep{Ananna2019} of all AGN. However, in the nearby Universe ($z <$ 0.1), CT-AGN represent only 5$-$10\% of the observed AGN \citep{Comastri_2004, Della_2008, Burlon2011, Ricci_2015, TorresAlba2021}.\\
\indent These sources are difficult to detect due to the significant obscuration of emission with energies $\leq$ 10\,keV \citep{Gilli2007, Koss2016}. Moreover, the majority of their emission comes from the so-called Compton hump at $\sim$20$-$40\,keV \citep{Gilli2007, Panagiotou2019}. Therefore, an instrument that is sensitive in this energy range is necessary to study CT-AGN in the local Universe. While the \textit{Swift}-Burst Alert Telescope (BAT) is capable of detecting these sources, it does not have the sensitivity required to accurately characterize CT-AGN \citep{Barthelmy2005}. Only the \textit{Nuclear Spectroscopic Telescope Array} \citep[\textit{NuSTAR};][]{Harrison_2013}, with a sensitivity two orders of magnitude greater than \textit{Swift}-BAT, can characterize the physical properties of these heavily obscured AGN \citep{Balokovic2014, Marchesi_2017b, Ursini2018, Zhao2019B, Zhao2019A, Balokovic2021}. However, AGN spectra at energies $\geq$ 10\,keV vary marginally with changing line-of-sight column density, whereas, soft X-rays ($<$ 10\,keV) vary significantly \citep[see, e.g.,][]{Gilli2007}. For that reason, \textit{XMM-Newton}, a soft X-ray instrument with the best effective area in 0.3$-$10\,keV ($\sim$10 times better than \textit{Swift}-XRT and $\sim$2 times better than \textit{Chandra}), is needed, in conjunction with \textit{NuSTAR}, to perform a robust characterization of obscured AGN. \\
\indent The Clemson-INAF Comtpon thick AGN project (CI-CTAGN)\footnote{\url{https://science.clemson.edu/ctagn/}} has been developed to find and characterize all obscured AGN in the local Universe by targeting CT-AGN candidates from the 150-month BAT catalog (Imam et al. in preparation). Our first step is to select high-latitude ($|b| >$ 10$\degr$), low-z ($z <$ 0.1) Seyfert 2 galaxies or sources classified as normal galaxies (as the absence of broad lines implies the presence of obscuring material in our line of sight) that do not have a ROSAT counterpart \citep{Voges1999} in the 0.1 $-$ 2.4\,keV band. Figure 2 from \cite{Koss2016} implies that any source at $z\approx0$ not detected by ROSAT will have a line-of-sight column density $\geq$ 10$^{23}$ cm$^{-2}$. Next, soft X-ray (\textit{Chandra}) snapshots ($\sim$10\,ks) are obtained and fit alongside BAT data to obtain preliminary column density measurements to identify the best obscured-AGN candidates \citep[see, e.g.,][hereafter, S22]{Marchesi17a, Marchesi_2017b, Silver2022}. The final step is to obtain simultaneous \xmm and \nus observations of these candidates to confirm their Compton-thick nature and to characterize the parameters of the torus, i.e., the obscuring dusty gas surrounding the SMBH. \\
\indent We have identified four nearby galaxies as obscured-AGN candidates, 2MASX J02051994$-$0233055 and 2MASX J04075215$-$6116126 from S22, and ESO 362$-$8 and IC 2227 from archival data. In this paper, we present the results of the \textit{NuSTAR}$-$\textit{XMM-Newton} analysis of these four sources. This work proceeds as follows: Section \ref{sec:obs} lists the observations and data reduction of our four sources. Section \ref{sec:spec_analysis} discusses the models used in analyzing the data and derived results. Section \ref{sec:disc} compares these new results to the previous values found in S22, as well as reports the progress our team has made thus far in detecting CT-AGN in the local Universe. All errors reported in this paper are at a 90\% confidence level. Standard cosmological parameters are as follows: H$_0$ = 70 km s$^{-1}$ Mpc$^{-1}$, q$_0$ = 0.0, and $\Lambda$ = 0.73.

\section{Observation and Data Analysis} \label{sec:obs}
The four sources we analyze are selected from the BAT 150-month catalog\footnote{\url{https://science.clemson.edu/ctagn/bat-150-month-catalog/}}, a catalog of 1390 AGNs that \bat detected in the 15--150\,keV band. Both 2MASX J02051994$-$0233055 and 2MASX J04075215$-$6116126 are listed as galaxies. Meanwhile, ESO 362$-$8 and IC 2227 are Seyfert 2 (Sy2) galaxies. \\
\indent 2MASX J02051994$-$0233055 was originally selected as a potentially heavily-obscured AGN in S22. 2MASX J04075215$-$6116126 was also studied in S22 and its selection is further discussed in Section \ref{sec:obs_2mj04}. Subsequently, they were targeted by \textit{Chandra} with 10\,ks snapshots (proposal ID 19700430, PI: Marchesi). The \textit{Chandra} data was fit with \bat to obtain a preliminary line-of-sight column density measurement for each source. 2MASX J02051994$-$0233055 had a best-fit N$_{\rm H, l.o.s.}$ $\sim$ 10$^{25}$ cm$^{-2}$ and 2MASX J04075215$-$6116126 had N$_{\rm H, l.o.s.}$ $\sim$ 2 $\times$ 10$^{23}$ cm$^{-2}$. The low statistics of \textit{Chandra} prevented us from confirming whether these sources were indeed CT-AGN and from characterizing properties of the torus. To do this, we obtained joint \textit{NuSTAR}$-$\textit{XMM-Newton} observations of each source (proposal ID 6220, PI: Ajello). ESO 362$-$8 and IC 2227 were selected as candidates following the procedure of S22, and existing archival data (\textit{XMM}; \textit{Swift}-XRT, respectively) were fit with BAT spectra, thus identifying them as CT-AGN candidates. Consequently, they were selected for joint \textit{NuSTAR}$-$\textit{XMM-Newton} observations as well (proposal ID 7219, PI: Silver). A summary of the observations is reported in Table \ref{tab:observations}.

\subsection{\textit{XMM-Newton} Observations}
All \xmm observations were reduced using the Science Analysis System \citep[sas,][]{Jansen2001} version 18.0.0. None of the observations were affected by flares. A 15$\arcsec$ circular region was used to extract the spectrum of each source. The background spectra were extracted using an annulus centered on the source with a 75$\arcsec$ inner radius and a 100$\arcsec$ outer radius. The image was visually inspected to ensure no contamination in the background from nearby sources. All three modules -- MOS1, MOS2, and pn -- are jointly fit in the modeling with their normalizations tied together\footnote{Our tests showed that leaving the 3 normalizations free to vary yields results consistent with those reported here.}, assuming marginal cross-calibration uncertainties. \\
\indent We note that spectra from the \textit{XMM-Newton}  Reflection Grating Spectrometer \citep[RGS,][]{denHerder2001} are available for the four sources, however they will not be analyzed in this work.

\subsection{\nus Observations}
\nus observed all sources quasi-simultaneously with \textit{XMM-Newton}, with the exception of 2MASX J04075215$-$6116126, as discussed below. The data is derived from both focal plane modules, FPMA and FPMB. The \texttt{nupipeline} version 0.4.8 was used to calibrate, clean, and screen the raw data files. The \nus calibration database (CALDB) version 20200813 was used in this analysis. The \texttt{nuproducts} script was used to produce the RMF, ARF, and light-curve files. For both modules, circular 50$\arcsec$ regions were used to extract the source spectra and an annulus with inner radius 100$\arcsec$ and outer radius 150$\arcsec$ were used to extract the background spectra. The images were visually inspected to verify no nearby sources contaminated the background. The HEAsoft task \texttt{grppha} was used to group both the \nus and \xmm data with 25 counts per bin. \\

\subsubsection{\nus Observations of 2MASX J04075215$-$6116126} \label{sec:obs_2mj04}
The first \nus observation of 2MASX J04075215$−$6116126, which was interrupted due to a ToO,  was originally centered on ESO 118$−$IG 004 NED01 as the target due to the mis-association of the BAT source and no significant X-ray emission was found at the center of the observation. We then discovered that 2MASX J04075215$−$6116126 was the true BAT counterpart by analyzing the \cha observation of this field in detail as presented in S22. Therefore, the following \nus observation (ID: 60601036002) was centered on 2MASX J04075215$−$6116126. Additionally, we note that the first \nus observation (60601027002) was taken near the South Atlantic Anomaly (SAA) and thus has higher background levels than typically encountered. Additionally, the true counterpart, 2MASX J04075215$-$6116126, was in the gap of the detector FPMB. For these reasons, this exposure could not provide valid scientific results and thus was not included in the analysis presented below. \\

\begin{deluxetable*}{cccccccc}
    \centering
    \tablecaption{Summary of \xmm and \nus Observations.}
    \label{tab:observations}
    \tablehead{\colhead{Source Name} & \colhead{Instrument} & \colhead{Sequence} & \colhead{Start Time} & \colhead{End Time} & \colhead{$z$} & \colhead{Exposure} & \colhead{Net Count Rate} \\
    \colhead{} & \colhead{} & \colhead{ObsID} & \colhead{(UTC)} & \colhead{(UTC)} & & \colhead{(ks)} & \colhead{10$^{-2}$ counts s$^{-1}$}}
    \startdata
    2MASX J02051994$-$0233055 & \xmm & 0870850101 & 2020-07-04 21:25 & 2020-07-05 07:32 & 0.0283 & 36.4 & 5.65 \\
    & \nus & 60601026002 & 2020-07-04 21:36 & 2020-07-05 06:10 & & 30.1 & 4.45 \\
    2MASX J04075215$-$6116126 & \xmm & 0870850201 & 2021-02-22 15:23 & 2021-02-23 15:18 & 0.0214 & 86.1 & 0.82 \\ 
    & \textit{NuSTAR} $^a$ & 60601027002 & 2021-02-22 14:26 & 2021-02-23 01:44 & & 40.7 & ... \\
    & \nus & 60601036002 & 2021-02-27 20:01 & 2021-02-28 08:56 & & 46.5 & 2.25 \\
    ESO 362$-$8 & \xmm & 0890440101 & 2021-10-05 13:10 & 2021-10-05 23:44 & 0.0158 & 38.0 & 0.26 \\
    & \nus & 60701048002 & 2021-10-05 02:11 & 2021-10-05 15:40 & & 48.5 & 1.41 \\
    IC 2227 & \xmm & 0890440201 & 2022-03-27 23:57 & 2022-03-28 10:31 & 0.0323 & 38.0 & 0.64 \\
    & \nus & 60701049002 & 2022-03-28 05:46 & 2022-03-28 20:15 & & 52.2 & 2.44 \\
    \enddata
\textbf{Notes:} \\
Average count rate (in cts s$^{-1}$), weighted by the exposure for \xmm
and \textit{NuSTAR}, where observations from multiple instruments are combined. Count rates are computed in the 2$-$10\,keV and 3$-$70\,keV band, respectively. \\
$^a$: This observation was not used in the analysis due to abnormally high background levels.
\end{deluxetable*}

\section{X-Ray Spectral Analysis} \label{sec:spec_analysis}
Spectral fitting was conducted with \texttt{XSPEC} v. 12.11.1 \citep{Arnaud1996}. The Galactic absorption in the direction of each source was calculated using the Heasoft tool \texttt{nh} \citep{Kalberla05}. \texttt{clumin}\footnote{\url{https://heasarc.gsfc.nasa.gov/xanadu/xspec/manual/node285.html}} in \texttt{xspec} was used to calculate the intrinsic luminosity of each source in the 2--10\,keV and 15--55\,keV bands. Tables \ref{tab:2m02nust}, \ref{tab:2m04nus}, \ref{tab:eso362}, and \ref{tab:ic2227} list the results of the 0.6$-$78\,keV spectral fitting. We note that each model implemented begins with a ``$constant_1$'' that accounts for flux variations between the \nus and \xmm observations. In this section, we introduce the physically-motivated models we used to fit the source spectra, in $\S$\ref{sec:models}, and the fitting results, in $\S$\ref{sec:results}.

\subsection{Models Implemented} \label{sec:models}
\subsubsection{\texttt{MYTorus}}
The first model applied in our analysis is \myt \citep{Murphy2009}. \myt assumes a torus of uniform absorbing material with circular cross section and an opening angle fixed to 60$\degree$, i.e., the covering factor = 0.50. \\
\indent The model is composed of three different components: an absorbed line-of-sight continuum (MYTZ), a Compton-scattered continuum (MYTS), and a fluorescent line emission (MYTL). These three components are linked together with the same normalization, absorbing column density (the equatorial column density of the torus, N$_{\rm H,eq}$), and inclination angle $\theta_i$. The inclination angle is measured from the axis of the torus, i.e., $\theta_i$=0$\degree$ represents a face-on view and $\theta_i$=90$\degree$ is an edge-on view. One can obtain the line-of-sight column density from the equatorial column density using

\begin{equation} \label{eq:nh,eq}
    N_{H,los} = N_{H,eq} \times (1 - 4 \times cos(\theta_i)^2)^{1/2}.
\end{equation}

The average torus column density is not a separate parameter as the model treats it as equal to the line-of-sight column density. However, it can be determined in certain configurations. \\
\indent The line-of-sight continuum, also called the zeroth-order continuum, is the intrinsic X-ray emission from the AGN observed after absorption from the torus along our line of sight. The Compton-scattered continuum is composed of the photons that interact with the dust and gas surrounding the SMBH and scatter into the observer line of sight. The final component includes the most significant fluorescent lines, i.e., the Fe K$\alpha$ and Fe K$\beta$, at 6.4 and 7.06\,keV, respectively. Both the reflected and fluorescent components are weighted by multiplicative constants, A$_S$ and A$_L$, respectively, that can account for differences in the geometry and time delays between the three components. Additionally, we include an additional component, $f_s$, to fit the fraction of intrinsic emission that escapes the torus instead of becoming absorbed. Lastly, our model includes \texttt{mekal} to account for the emission below 3\,keV caused by diffuse hot gas. \myt can be used in either the `coupled' or `decoupled' configuration (see $\S$\ref{sec:mytdec}). The model in \texttt{XSPEC} notation is as follows:

\begin{eqnarray}
 \label{eq:myt}
    ModelA = constant_1  * phabs * \nonumber (MYTZ * \\ zpowerlw + \nonumber A_S * MYTS + \nonumber A_L * MYTL \nonumber \\ + f_s * zpowerlw + mekal).
\end{eqnarray}

In this work, we only present results using the decoupled configuration as \texttt{MYTorus} coupled has been shown to yield statistically-worse fits and provides less information about the obscuring material average properties \citep[see e.g.,][]{TorresAlba2021}.

\subsubsection{\texttt{MYTorus} in Decoupled Configuration} \label{sec:mytdec}
Unlike the coupled configuration, the \myt decoupled configuration \citep{Yaqoob2012} allows for the separate measurement of the line-of-sight column density, N$_{H, l.o.s.}$, and the average torus column density, N$_{H, avg}$, thus mimicking a clumpy torus distribution. In this arrangement, the line-of-sight continuum has a fixed inclination angle of $\theta_{i,Z}$=90$\degree$. The reflected and fluorescent line components can have their inclination angles fixed to both $\theta_{i,S}$=90$\degree$ and $\theta_{i,S}$=0$\degree$, representing an edge-on or face-on scenario, respectively. Additionally, their column densities are tied together to the average torus column density N$_{H, avg}$. \\


\subsubsection{BORUS02}
The next physically motivated model utilized in this work is \texttt{borus02} \citep{Balo_borus2018}. Like \texttt{MYTorus}, \bor assumes a uniform obscuring material, however, the opening angle is not fixed. Thus, the covering factor $c_f$ is a free parameter ($c_f \in$ [0.1,1]). The model only contains a reflection component, which includes both the reflection continuum and fluorescent lines. Therefore, we manually add the absorbed intrinsic continuum multiplied by a line-of-sight absorbing component, \texttt{zphabs} $\times$ \texttt{cabs}. \bor is implemented in \texttt{XSPEC} as follows:

\begin{equation}
\begin{aligned}
 \label{eq:borus}
    ModelB = constant_1  * phabs * \\ (borus02 + zphabs * cabs * zpowerlw \\
    + f_s * zpowerlaw).
\end{aligned}
\end{equation}

Similarly to the decoupled configuration of \texttt{MYTorus}, \bor is capable of measuring both the line-of-sight and average torus column density. However, unlike \myt decoupled, \bor can constrain the observing angle in the range cos($\theta_{inc}$) = 0.05--0.9. \\
\indent \bor also includes a high-energy cutoff which we freeze to 500\,keV. We note that recent works find a lower average cutoff energy \citep[$\sim$ 200--300\,keV;][]{Ricci2017, Ananna_2020, Balokovic2021}. However, the \nus data for our sources corresponds to $<$80\,keV in the source rest-frame, thus this change in high-energy cutoff would not affect our results.

\subsubsection{UXClumpy}
Unlike \bor and \texttt{MYTorus}, \uxc \citep{Buchner2019} does not assume a uniform torus. Instead, \uxc is a physically motivated model that reproduces the data by simulating different cloud sizes and distributions. \uxc utilizes a Monte Carlo X-ray radiative transfer code, XARS, to compute the X-ray spectra of obscured AGN. The model is implemented in \texttt{XSPEC} as follows:

\begin{eqnarray}
 \label{eq:myt}
    ModelC = constant_1  * \nonumber phabs * \\  \nonumber (uxcl\_cutoff.fits + \\ f_s * uxcl\_cutoff\_omni.fits).
\end{eqnarray}

\noindent The first table accounts for the transmitted and reflection components, including fluorescent lines. \uxc produces the reflection component through the cloud distribution it generates. However, for some sources that are reflection-dominated, a Compton-thick reflector near the corona can be added. This can be thought of as an inner wall that blocks the line of sight to the corona while also reflecting its emission. The second table reproduces the intrinsic continuum that leaks through the clumps of the torus. \\
\indent \uxc differs from \bor in that it does not include a parameter to measure the average torus column density. However, it measures other torus parameters such as the inclination angle (with a slightly larger range than \texttt{borus02}; cos($\theta_{inc}$) = 0--1.00), the dispersion of the cloud distribution \texttt{TORsigma} ($\sigma$ ranges from 6--90$\degr$), and the covering factor of the inner reflector \texttt{CTKcover} (C ranges from 0--0.6).

\subsection{Fitting Results} \label{sec:results}

The spectra and resulting best-fit parameters can be found in Figures \ref{fig:2m02}, \ref{fig:2m04}, \ref{fig:eso362}, \ref{fig:ic22} and Tables \ref{tab:2m02nust}, \ref{tab:2m04nus}, \ref{tab:eso362}, \ref{tab:ic2227}, respectively. We note that the spectra of every source were fit starting from 0.6\,keV, as this is the minimum allowed energy in \texttt{MYTorus}. We kept the same value in \bor and \uxc for consistency. Additionally, we left the \nus cross-normalization constant $c_{nus}$ free to vary in all models as even quasi-simultaneous \xmm and \nus observations can differ in the measured flux by up to 10\% \citep[see Table 5,][]{Madsen2017}.

\subsubsection{2MASX J02051994$-$0233055}
Unlike the initial fits of 2MASX J02051994$-$0233055 in S22 which suggested it was a CT-AGN with line-of-sight column density $\sim$ 10$^{25}$ cm$^{-2}$, the \nus and \xmm data are consistent with a power law. Therefore, we fit the data as such:

\begin{equation}
\label{eq:plaw}
    ModelD = constant_1 * phabs * (zphabs * zpowerlw).
\end{equation}

The results of this fit are presented in Table \ref{tab:2m02nust} and the spectra in Figure \ref{fig:2m02}. The best-fit result for the line-of-sight column density is on the order of 10$^{20}$ cm$^{-2}$, consistent with an unobscured AGN. Moreover, adding a reflection component does not statistically improve the fit. We will further discuss this significant discrepancy in Section \ref{sec:comp_results}.

\subsubsection{2MASX J04075215$-$6116126}
The best-fit results in Table \ref{tab:2m04nus} show relatively consistent values between the different models. For example, all models yield a line-of-sight column density $\approx$ 0.30 $\times$ 10$^{24}$ cm$^{-2}$. Furthermore, all models agree that the source is observed through a less dense portion of the torus as the average column density has a larger value, even entering into the Compton-thick regime in the \bor best-fit results. Additionally, all models yield a photon index $\Gamma \sim$ 1.48, a best-fit value somehow harder than it is commonly measured in AGN. To test the validity of this result, we froze the photon index to 1.8 and refit yielding largely similar results and near indistinguishable fit statistics \citep[a similar method was originally established in][]{Nandra1994}. As a consequence, we are unable to say which photon index is a better physical representation of this source. This is consistent with the large uncertainties measured. \\
\indent Despite these observations taking place only five days apart, we find significant flux variation between \nus and \xmm with a best-fit \nus cross-normalization constant $\approx$ 1.45. To verify this variation, we included \textit{Chandra} data from 2018 (Obs ID: 20440, exposure: 10.4\,ks) and \textit{Swift}-XRT data from 2018 (Obs ID: 00094011001, exposure 2.3\,ks) and 2021 (Obs ID: 00089208002, exposure 2.1\,ks). The two 2018 observations had a cross-normalization constant greater than 3 and the 2021 XRT observation had a constant $\approx$ 2. Furthermore, we tested if these variations could instead be coming from a column density fluctuation rather than intrinsic luminosity. When we left all the cross-normalization constants fixed to one and decoupled the column density of each observation, we found all observations yield a consistent line-of-sight N$_{\rm H, l.o.s.}$ value $\approx$ 0.30 $\times$ 10$^{24}$ cm$^{-2}$. Therefore, we confirm that 2MASX J04075215$-$6116126 experienced a nearly 50\% flux variation in just five days time and significant variation over two years.

\subsubsection{ESO 362$-$8}
The initial fits of ESO 362$-$8 featured a very soft photon index ($\Gamma$ = 2.6 in \texttt{borus02}) which is atypical of AGN, whose average value lies around $\sim$1.6--1.8 \citep{Ricci2017}. This might be caused by an unusual excess in soft emission. To account for this, we first added a second \texttt{mekal} component. However, this only reduced the photon index to 2.4. Our next test decoupled the scattering emission photon index from the intrinsic emission photon index \citep[as used in][]{Nuria2018}. This difference in photon index stems from the contribution of X-ray binaries in sources with significant star formation. \nus has recently been used to properly model these luminous and ultraluminous infrared galaxies \citep[U/LIRGs;][]{Teng2015, Puccetti2016, Ricci2021, Yamada2021}.
We find this to be our most physically plausible representation of the data, as the main power law has values from 1.70--1.90, and the scattered power law accounts for the soft excess with values from 2.90--3.00. \\
\indent All models do agree that ESO 362$-$8 is a bona-fide Compton-thick AGN, with line-of-sight N$_{\rm H, l.o.s.}$ ranging from ~2--4$\times$ 10$^{24}$ cm$^{-2}$. This is the first, and only, confirmed Compton-thick AGN in this paper. Most applicable models also agree that the torus is Compton-thick with N$_{\rm H, avg}$ $\sim$ 1$\times$ 10$^{25}$ cm$^{-2}$, however the \myt edge-on fit only has an N$_{\rm H, avg}$ = 1.5$\times$ 10$^{23}$ cm$^{-2}$. This discrepancy may be caused by the fact that we are viewing the source nearly face-on (as supported by \bor and \texttt{UXClumpy}), while this model configuration tries to force the edge-on view. Finally, \bor and \uxc agree on the parameters constraining the torus, such as the near face-on inclination angle ($\sim$0.90--1.00) and a significant covering factor (0.90 for \bor and 0.31 for \texttt{UXClumpy}).

\subsubsection{IC 2227}
All models agree that IC 2227 is a Compton-thin AGN with line-of-sight N$_{\rm H, l.o.s.}$ $\sim$ 0.6$\times$ 10$^{24}$ cm$^{-2}$. They are also consistent with yielding a photon index around 1.8. Furthermore, the models agree that this source is reflection dominated due to its Compton-thick average torus N$_{\rm H, avg}$, ranging from 1.4--31$\times$ 10$^{24}$ cm$^{-2}$, and a large covering factor of 0.80 and 0.60 from \bor and \texttt{UXClumpy}, respectively. 

\section{Discussion} \label{sec:disc}
This work serves as the third step in our previously proven successful process \citep{Zhao2019B, Zhao2019A} of identifying and characterizing CT-AGN in the local Universe. First, we used the selection criteria laid out in S22 to discover potentially obscured AGN and propose them to \textit{Chandra}. Next, we analyze the \textit{Chandra} snapshots along with \textit{Swift}-BAT data to determine a preliminary line-of-sight column density value. Finally, we use these results to pick the best CT-AGN candidates and propose for joint \textit{NuSTAR}$-$\textit{XMM-Newton} observations, thus allowing us to confirm whether or not these candidates are Compton thick and to measure properties of the obscuring material, such as its average column density and covering factor.

\subsection{Comparison with Previous Results} \label{sec:comp_results}

\subsubsection{2MASX J02051994$-$0233055} \label{sec:2m_prev_res}
S22 listed 2MASX J02051994$-$0233055 as a Compton-thick candidate, finding a line-of-sight column density of 1 $\times$ 10$^{25}$ cm$^{-2}$, however with large uncertainties ($\sim$70\%). The \textit{NuSTAR}$-$\textit{XMM} analysis discovered that 2MASX J02051994$-$0233055 is not a CT-AGN, in fact, it is an unobscured AGN. Our typical obscured-AGN models described in Section \ref{sec:models}, which include significant contribution from a reprocessing component, were unable to satisfactorily fit the data. Instead, an absorbed power law was used and found a line-of-sight column density of 3 $\times$ 10$^{20}$ cm$^{-2}$ with smaller uncertainties ($\sim$30\%). \\
\indent The \textit{Chandra}$-$BAT analysis labeled this as a CT-AGN candidate likely due to the source being in an extremely low flux state during the \textit{Chandra} observation. To confirm this variability, we plotted in Figure \ref{fig:2m02alld} the \textit{XMM-Newton} (magenta, orange, and yellow) and \textit{NuSTAR} (blue and cyan) data alongside the BAT (red), \textit{Chandra} observation (from June 2018, black) and \textit{Swift}-XRT observation (from June 2018, green). The XRT observation, taken during the same week as the \textit{Chandra} observation, has a similar flux level (see Figure \ref{fig:2m02_flux}), confirming the variability. Even more interestingly, the BAT data (which is an average over 150 months), is at a higher flux level\footnote{The 2--10\,keV BAT flux was extrapolated by fitting the BAT spectra using a power law with the photon index frozen to the best-fit value from fitting the soft X-ray data and assuming the same obscuration (see Table \ref{tab:2m02nust}). To calculate the errors, we repeated the procedure at the upper and lower errors of the normalization.} than even the \textit{NuSTAR} and \textit{XMM} data. This suggests that if the \textit{Chandra} state has a flux 5$\times$ lower than the BAT data (in the 2--10\,keV band), there could also have been a time when the source was in a flux state 5$\times$ higher than the BAT data. \\ 
\indent We note that while some AGN have shown line of sight N$_{\rm H}$ variability \citep[see e.g.,][]{Risaliti2010, Markowitz2014, Laha2020, Pizzetti2022}, no source has yet varied from a Compton thick AGN state to an unobscured one. Therefore, it is much more likely that intrinsic luminosity variability is responsible for the change in spectral shape of 2MASX J02051994$-$0233055, as is supported statistically by our fits. This source marks the first time since beginning our search for CT-AGN that our selection criteria yielded an unobscured AGN. Such a result further highlights the importance of simultaneous \nus and \textit{XMM-Newton} observations in determining the column density of AGN. \\

\subsubsection{2MASX J04075215$-$6116126}
\indent The joint \textit{Chandra}-\textit{Swift}-BAT spectrum of 2MASX J04075215$-$6116126 was analyzed in S22 and found to be a Compton-thin candidate, with line-of-sight column density of 2.10$^{+0.04}_{-0.08}$ $\times$ 10$^{23}$ cm$^{-2}$ (this is the \bor result; other models produced similar values). The \textit{NuSTAR}$-$\textit{XMM-Newton} analysis presented in this paper yielded similar results ($\sim$3 $\times$ 10$^{23}$ cm$^{-2}$), confirming this source to be a Compton-thin AGN. This work found the average torus column density to be larger than previously found, even entering the Compton-thick regime in the \bor results. However, this difference could be caused by the larger uncertainties from the \textit{Chandra}$-$BAT fits ($>$140\% uncertainties versus $\sim$80\% uncertainties in this work). \\

\subsubsection{ESO 362$-$8}
\indent Neither ESO 362$-$8 nor IC 2227 have previously published N$_{\rm H}$ values. Instead, we compare to the results found from fitting the archival data with BAT spectra. We note that at the time of the proposal, neither source had BAT spectra available to us. Instead, the archival data was jointly fit with BAT data from other sources that were newly discovered in the 150-month catalog (just as ESO 362$-$8 and IC 2227 were, and thus are expected to have very similar flux levels). \\
\indent The 18\,ks archival \xmm observation of ESO 362$-$8 from February 2006 yielded a photon index of 1.73 and an N$_{\rm H, l.o.s.}$ = 1.25 $\times$ 10$^{24}$ cm$^{-2}$. The photon index is in good agreement with the simultaneous \nus and \textit{XMM} data, as most models yielded $\sim$1.8. The new results also confirmed this source as Compton-thick, however with a larger N$_{\rm H, l.o.s.}$, $>$2 $\times$ 10$^{24}$ cm$^{-2}$, than found in the archival data. \\

\subsubsection{IC 2227}
\indent The 20\,ks archival XRT data from May 2008 for IC 2227 produced a best fit photon index of 1.81 and N$_{\rm H, l.o.s.}$ = 1.23 $\times$ 10$^{24}$ cm$^{-2}$. The \nus and \xmm data found a similar photon index, with most models around 1.85. However, the new data found IC 2227 to be Compton-thin, not Compton-thick as predicted by the XRT results. While the archival data is consistent with a Compton-thin scenario within 90\% confidence (9 $\times$ 10$^{23}$ cm$^{-2}$, see the blue line in Figure \ref{fig:ic22}), it does not fall as low as the 6 $\times$ 10$^{23}$ cm$^{-2}$ value found by the new data. There are at least two possible explanations to this discrepancy: 1) The XRT data was not of a high enough quality to accurately estimate the true N$_{\rm H, l.o.s.}$ of the source \citep[][found that XRT+BAT fits often over-estimate N$_{\rm H, l.o.s.}$]{Marchesi2018}, or 2) IC 2227 experienced variability in its line-of-sight column density between the XRT observation in 2008-2009 and its \nus and \xmm observations in 2022. This source may be targeted again in the future to identify if there is true N$_{\rm H, l.o.s.}$ variability present.

\subsection{Clemson-INAF CT-AGN Project}
Using joint fits of soft X-ray and BAT data, \cite{Ricci_2015} presented a list of CT-AGN candidates in the 70-month BAT catalog. 55 sources were listed, with 50 having $z \leq 0.05$. Adding sources from the Palermo 100 catalog \citep{Cusumano2014}, more recent works \citep{Marchesi17a, Marchesi_2017b}, and the four sources presented in this paper, brings this list up to 65 CT-AGN candidates with $z \leq 0.05$. Including this work, our group has now personally analyzed 52 of these sources, confirming 28 to be CT-AGN based on their simultaneous \textit{NuSTAR}--\textit{XMM} data. This is a roughly $\sim$50\% success rate, highlighting the significance of \nus for confirming sources as CT-AGN. In total, there have now been 35 CT-AGN discovered in the local Universe\footnote{\url{https://science.clemson.edu/ctagn/ctagn/}} \citep{TorresAlba2021}.

\subsection{Observational evidence for non-homogeneity of the obscuring material}
\begin{figure}
    \hspace{-1.cm}
    \includegraphics[scale=0.65, clip=true,trim=0 0 0 0mm]{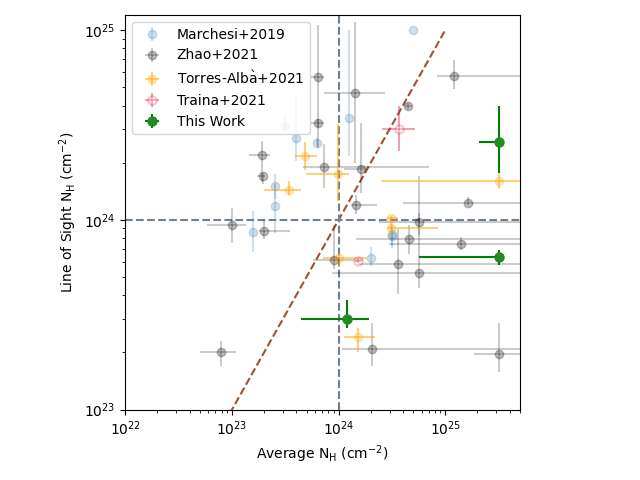}
    \caption{The \bor best fit values of line-of-sight column density versus the average column density of the AGN in this project. The three sources in this work are shown in green. 2MASX J02051994$-$0233055 is not included as it is unobscured, and thus, we are unable to provide an average torus column density measurement. The vertical and horizontal dashed lines represent the CT threshold, while the diagonal dashed line is the one-to-one relationship between N$_{\rm H,los}$ and N$_{\rm H,avg}$. Other sources are from \cite{Marchesi_2019, TorresAlba2021, Traina2021, Zhao2021}.}
    \label{fig:nh_compare}
\end{figure}

\begin{figure*}
    \centering
    \includegraphics[scale=0.75]{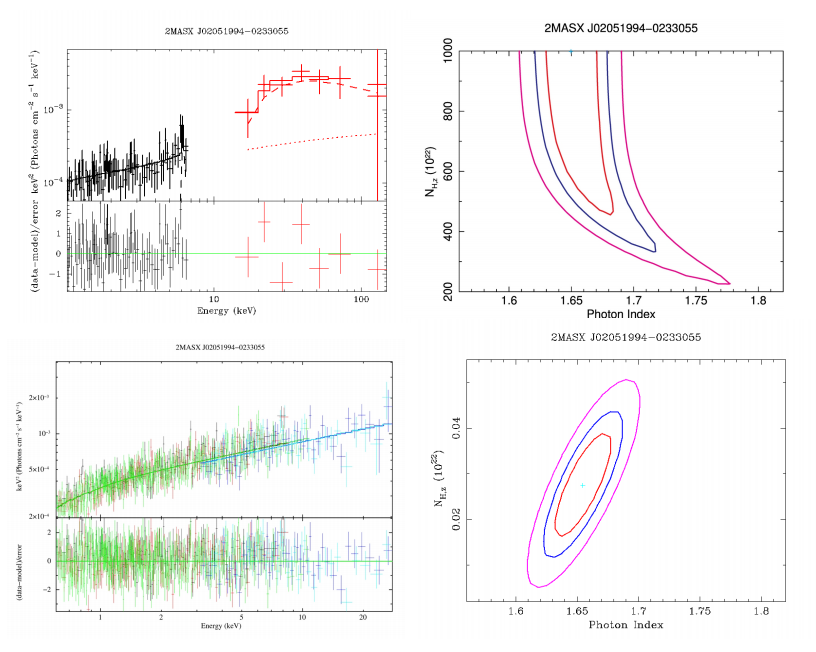}
    \caption{Top: The \bor fit of the \textit{Chandra}-BAT data presented in S22 with the contours of N$_{H, l.o.s.}$ ($\times$ 10$^{22}$ cm$^{-2}$) vs Photon Index. The contours represent the 68, 90 and 99\% confidence levels. Bottom: The \textit{NuSTAR}$-$\textit{XMM-Newton} data fit with a power law alongside its contour of the same parameters. }
    \label{fig:2m02}
\end{figure*}

\begin{figure*}
    \centering
    \includegraphics[scale=0.75]{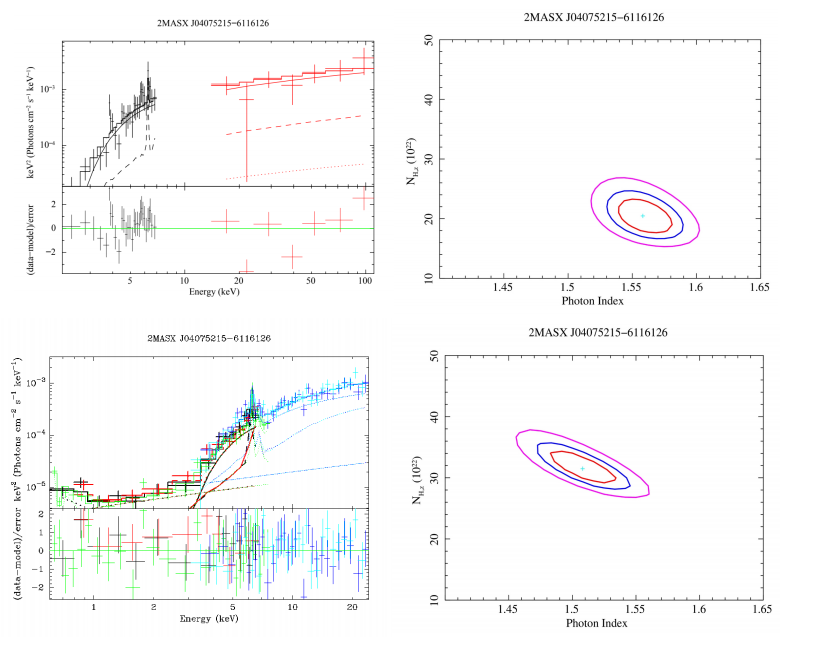}
    \caption{Top: The \bor fit for the \textit{Chandra}-BAT data of 2MASX J04075215$-$6116126 presented in S22 with the contours of N$_{H, l.o.s.}$ ($\times$ 10$^{22}$ cm$^{-2}$) vs Photon Index. The contours represent the 68, 90 and 99\% confidence levels. Bottom: The \bor fit of the \textit{NuSTAR}$-$\textit{XMM-Newton} data alongside its contour of the same parameters. }
    \label{fig:2m04}
\end{figure*}

\begin{figure*}
    \centering
    \includegraphics[scale=0.75]{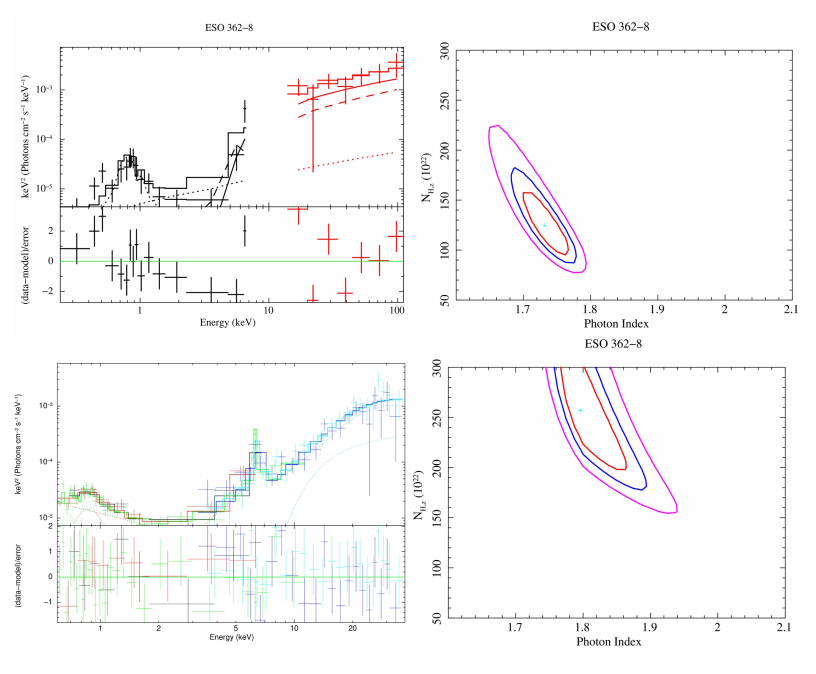}
    \caption{Top: The \bor fit for the archival \textit{XMM}-BAT data of ESO 362$-$8 with the contours of N$_{H, l.o.s.}$ ($\times$ 10$^{22}$ cm$^{-2}$) vs Photon Index. The contours represent the 68, 90 and 99\% confidence levels. Bottom: The \bor fit of the \textit{NuSTAR}$-$\textit{XMM-Newton} data alongside its contour of the same parameters. }
    \label{fig:eso362}
\end{figure*}

\begin{figure*}
    \centering
    \includegraphics[scale=0.75]{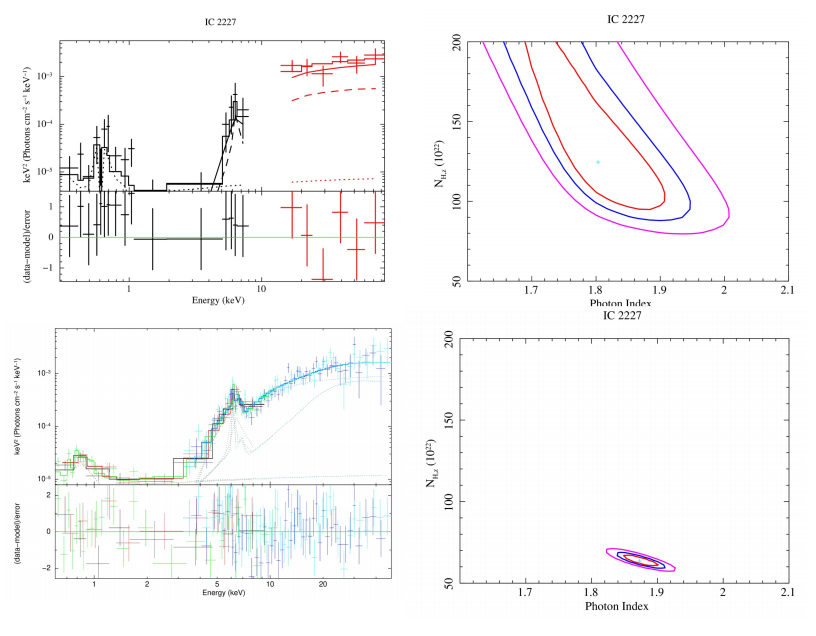}
    \caption{Top: The \bor fit for the archival XRT-BAT data of IC 2227 with the contour of N$_{H, l.o.s.}$ ($\times$ 10$^{22}$ cm$^{-2}$) vs Photon Index. The contours represent the 68, 90 and 99\% confidence levels. Bottom: The \bor fit of the \textit{NuSTAR}$-$\textit{XMM-Newton} data alongside its contour of the same parameters. }
    \label{fig:ic22}
\end{figure*}

\begin{figure*}
    \centering
    \includegraphics[scale=0.1, angle=270]{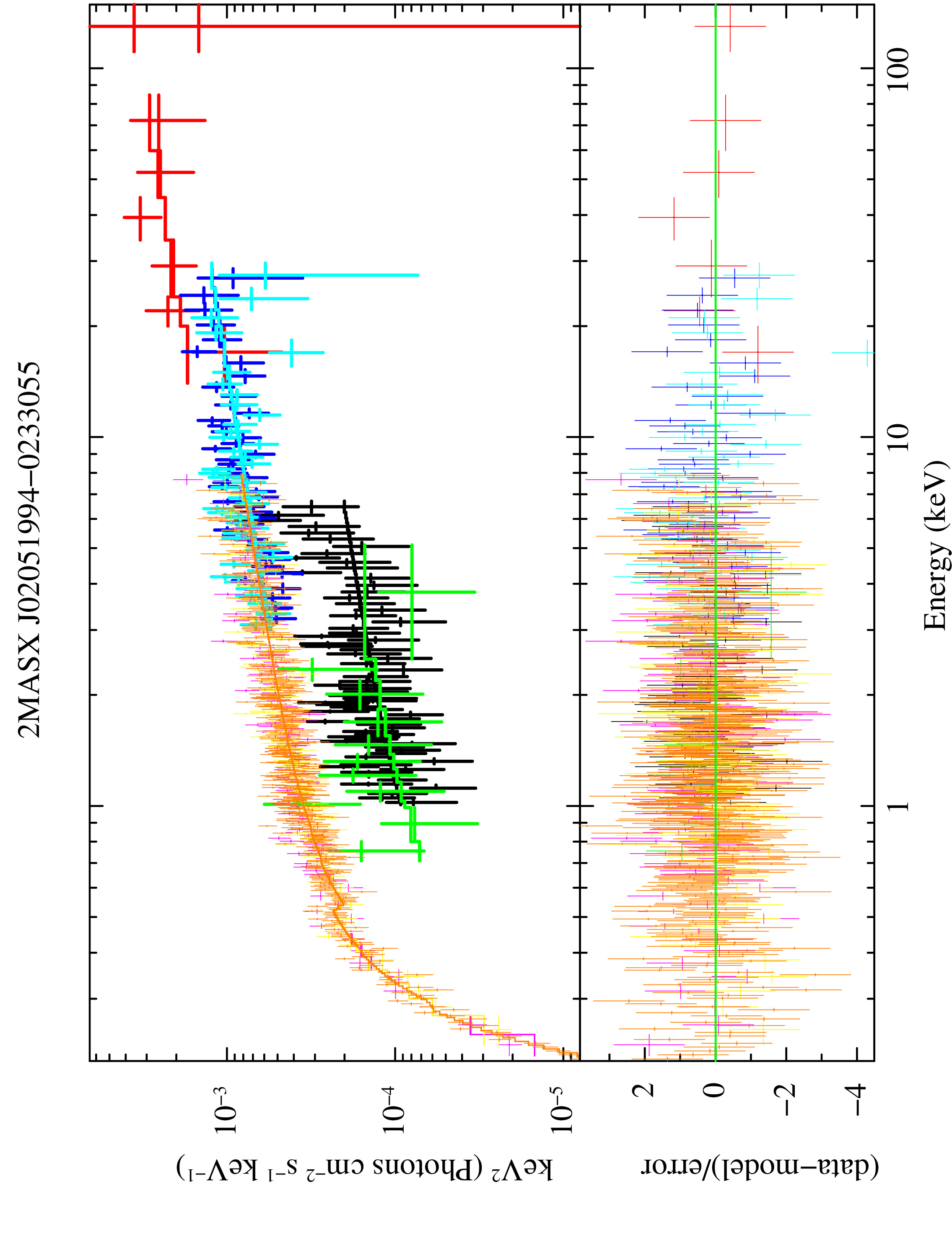}
    \caption{Data of 2MASX J02051994$-$0233055 from multiple instruments demonstrating the source's flux variability while maintaining a consistent photon index. The \textit{Chandra} data is in black (taken in 2018); BAT (average spectrum obtained combining the data taken between 2005 and 2017) in red; \textit{XMM-Newton} in orange, yellow, and magenta (2021); \textit{NuSTAR} in blue and cyan (2021); \textit{Swift}-XRT in green (2018).}
    \label{fig:2m02alld}
\end{figure*}

\begin{figure*}
    \centering
    \includegraphics{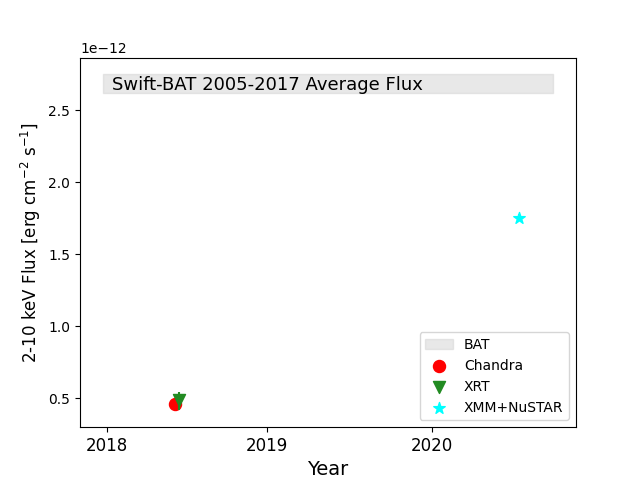}
    \caption{The 2$−$10\,keV flux of 2MASX J02051994$−$0233055 as a function of time over the past 15 years. The red circle and green triangle are the best-fitted fluxes from Chandra (2018 June 11 Date) and XRT (2008 June 25) observations. The cyan star is the source flux obtained in July 2020 by \nus and \textit{XMM-Newton}. The grey filled region shows the 12 years average flux from the Swift-BAT observations in 2005-2017, which is converted from its 14-195\,keV flux to  2$-$10\,keV flux. The light curve suggests that 2MASX J02051994$−$0233055 has experienced a more than 5 times flux variability in the last 15 years due to the different accretion rates as analyzed in Section \ref{sec:2m_prev_res}. Uncertainties on the fluxes are plotted but are too small to be visible.}
    \label{fig:2m02_flux}
\end{figure*}

Figure \ref{fig:nh_compare} compares the line-of-sight column density with the average torus column density of CT-AGN candidates studied as a part of this project \citep[see, ][]{Marchesi_2019, TorresAlba2021, Traina2021, Zhao2021}. The figure shows no visible trend between the two values, i.e., Compton-thick AGN are no more likely to have Compton-thick tori compared to less obscured AGN. This supports the idea that the material causing the X-ray obscuration is not a homogeneous structure. Instead, it is comprised of differing density clumps that revolve around the central engine, moving into and out of our line of sight. This can lead to different N$_{\rm H, l.o.s.}$ measurements when a source has multi-epoch observations. This has been proven in recent works on sources such as NGC 7479 \citep{Pizzetti2022}, NGC 1358 (Marchesi et al. accepted), and in a sample of Compton-thin \citep{Zhao2021} and Compton-thick (Torres-Alb\`{a} in prep.) sources. The three obscured AGN in our sample\footnote{Since 2MASX J02051994$-$0233055 is found to be unobscured, its reprocessed emission cannot be reliably measured. As a consequence, no measurement of the average torus column density can also be performed.} (i.e., excluding 2MASX J02051994$-$0233055) all lie away from the diagonal dashed line, thus further supporting this hypothesis. As already discussed, this difference is especially true with IC 2227, making it a potential candidate for future monitoring.

\section{Conclusions}
In this work, we have analyzed simultaneous \nus and \xmm data of 4 CT-AGN candidates with the physically motivated tori models \texttt{MYTorus}, \texttt{borus02}, and \texttt{UXClumpy}. None of the sources have had \nus data published previously. We summarize our conclusions as follows:

\begin{itemize}
    \item Of the 4 sources analyzed, one, ESO 362$-$8, is confirmed to be a bona-fide CT-AGN. This increases the sample of BAT-detected CT-AGN in the local Universe (z $<$ 0.1) to 35. 
    \item 2MASX J02051994$-$0233055 was determined to be a highly flux-variable, unobscured AGN. This is the first source studied using our criteria to select heavily obscured AGN that was instead discovered to be unobscured due to its strong flux variability. This highlights the importance of simultaneous soft and hard X-ray observations to accurately classify and characterize the heavily obscured AGN population. 
    \item 2MASX J04075215$-$6116126 displayed significant flux variation ($\sim$50\%) in only five days separating observations. Moreover, the flux varied by a factor of 3 when compared with observations taken 2 years prior. It was confirmed that N$_{\rm H,los}$ remained constant during these periods, thus providing another example of an AGN with significant luminosity variation. Such sources can be studied to probe the relationship between the luminosity and the geometry of the obscuring (i.e., covering factor).
    \item All three sources with N$_{\rm H,los}$ $>$ 10$^{23}$ cm$^{-2}$  show statistically significant differences in their line-of-sight and average torus column densities. This further supports that the structure of the obscuring material surrounding accreting SMBHs may be clumpy, rather than uniform. 
\end{itemize}

\acknowledgements
The authors thank the anonymous referee for their detailed and helpful comments which greatly improved the paper. \\
\indent RS, NTA, AP, and MA acknowledge NASA funding under contracts 80NSSC20K0045, 80NSSC19K0531, and 80NSSC21K0016 and SAO funding under contracts GO0-21083X and G08-19083X. SM acknowledges funding from the the INAF ``Progetti di Ricerca di Rilevante Interesse Nazionale'' (PRIN), Bando 2019 (project: ``Piercing through the clouds: a multiwavelength study of obscured accretion in nearby supermassive black holes''). 

\begin{deluxetable*}{ccccccccc}

\tablecaption{Power law fit of 2MASX J02051994$-$0233055 \nus-\xmm}
\label{tab:2m02nust}

\tablehead{\colhead{$\chi^2$/dof} & \colhead{$\Gamma$} & \colhead{$N_{H,los}$} & \colhead{norm} & \colhead{$c_{nus}$} & \colhead{Flux} & \colhead{Flux} & \colhead{Lum.} & \colhead{Lum.} \\ \colhead{} & \colhead{} & \colhead{} & \colhead{$10^{-2}$} & \colhead{} & \colhead{$\rm_{2-10\,keV}$} & \colhead{$\rm_{15-55\,keV}$} & \colhead{$\rm_{2-10\,keV}$} & \colhead{$\rm_{15-55\,keV}$}}

\startdata
839/862 & 1.64$^{+0.02}_{-0.02}$ & 0.0003$^{+0.0001}_{-0.0001}$ & 0.041$^{+0.001}_{-0.001}$ & 0.96$^{+0.05}_{-0.05}$ & 1.75$^{+0.03}_{-0.02}$ $\times$ 10$^{-12}$ & 2.62$^{+0.06}_{-0.10}$ $\times$ 10$^{-12}$ & 3.16$^{+0.23}_{-0.14}$ $\times$ 10$^{42}$ & 5.01$^{+0.24}_{-0.33}$ $\times$ 10$^{42}$ \\
\enddata
\vspace{5mm}
\textbf{Notes:} \\ 
$\chi^2$/dof: $\chi^2$ divided by degrees of freedom. \\
$\Gamma$: Power law photon index. \\
$N_{H, l.o.s.}$: line-of-sight torus hydrogen column density, in units of 10$^{24}$ cm$^{−2}$. \\
norm: the main power-law normalization (in units of photons cm$^2$ s$^{−1}$ keV$^{−1}$ $\times$ 10$^{−4}$), measured at 1 keV. \\
$c_{nus}$: The cross-normalization constant between the \textit{XMM} and \textit{NuSTAR} data. \\
F$\rm_{2-10\,keV}$: Observed flux in the 2$-$10\,keV band with units of erg cm$^{-2}$ s$^{-1}$. \\
F$\rm_{15-55\,keV}$: Observed flux in the 15$-$55\,keV band with units of erg cm$^{-2}$ s$^{-1}$. \\
L$\rm_{2-10\,keV}$: Intrinsic luminosity in the 2$-$10\,keV band with units of erg s$^{-1}$. \\
L$\rm_{15-55\,keV}$: Intrinsic luminosity in the 15$-$55\,keV band with units of erg s$^{-1}$. \\
\end{deluxetable*}

\newpage

\begin{deluxetable*}{ccccccc}[h!]

\tablecaption{2MASX J04075215$-$6116126 \nus - \xmm}
\label{tab:2m04nus}

\tablehead{\colhead{Model} & \colhead{MYTorus} & \colhead{MYTorus} & \colhead{borus02} & \colhead{borus02} & \colhead{UXClumpy }\\ 
\colhead{} & \colhead{(Decoupled Face-on)} & \colhead{(Decoupled Edge-on)} & \colhead{} & \colhead{$\Gamma$=1.80}}

\startdata
$\chi^2$/dof & 115/116 & 126/116 & 110/114 & 112/115 & 121/115 \\
kT & 0.23$^{+0.08}_{-0.08}$ & 0.23$^{+0.08}_{-0.08}$ & 0.25$^{+0.09}_{-0.07}$ & 0.23$^{+0.09}_{-0.09}$ & 0.24$^{+0.16}_{-0.13}$ \\
$\Gamma$ & 1.48$^{+0.22}_{-u}$ & 1.47$^{+0.18}_{-u}$ & 1.49$^{+0.36}_{-u}$ & 1.80* & 1.73$^{+0.25}_{-0.13}$ \\
norm $10^{-2}$ & 0.01$^{+0.01}_{-0.01}$ & 0.02$^{+0.01}_{-0.01}$ & 0.01$^{+0.02}_{-0.01}$ & 0.03$^{+0.01}_{-0.01}$ & 0.03$^{+0.27}_{-0.01}$ \\
c$_{f,Tor}$ & ... & ... & 0.60$^{+u}_{-0.33}$ & 0.54$^{+0.37}_{-0.24}$ & ... \\
CTKcover & ... & ... & ... & ... & 0* \\
Tor $\sigma$ & ... & ... & ... & ... & 28.2$^{+u}_{-18.2}$ \\
cos($\theta_{obs}$) & ... & ... & 0.85$^{+u}_{-0.38}$ & 0.75$^{+u}_{-0.21}$ & 1.00$^{+u}_{-u}$ \\
$N_{H, l.o.s.}$ & 0.31$^{+0.05}_{-0.03}$ & 0.31$^{+0.06}_{-0.04}$ & 0.30$^{+0.08}_{-0.03}$ & 0.36$^{+0.06}_{-0.05}$ & 0.33$^{+0.08}_{-0.04}$ \\
$N_{H, avg}$ & 0.80$^{+0.51}_{-0.41}$ & 0.40$^{+0.44}_{-0.22}$ & 1.20$^{+0.71}_{-0.76}$ & 5.75$^{+u}_{-4.49}$ & ... \\
$f_s$ 10$^{-2}$ & 3.56$^{+1.10}_{-1.40}$ & 3.43$^{+1.00}_{-1.10}$ &  3.61$^{+0.80}_{-2.30}$ & 1.89$^{+0.69}_{-0.44}$ & 8.24$^{+13.0}_{-u}$\\
c$_{nus}$ & 1.45$^{+0.16}_{-0.14}$ & 1.44$^{+0.17}_{-0.16}$ & 1.50$^{+0.17}_{-0.14}$ & 1.57$^{+0.15}_{-0.16}$ & 1.55$^{+0.16}_{-0.16}$ \\
\hline
F$\rm_{2-10\,keV}$ & 2.60$^{+0.42}_{-1.06}$ $\times$ 10$^{-13}$ & 2.57$^{+0.46}_{-1.12}$ $\times$ 10$^{-13}$ & 2.56$^{+0.39}_{-1.11}$ $\times$ 10$^{-13}$ & 2.55$^{+0.12}_{-0.53}$ $\times$ 10$^{-13}$ & 2.58$^{+0.23}_{-2.89}$ $\times$ 10$^{-13}$ \\
F$\rm_{15-55\,keV}$ & 2.23$^{+0.43}_{-0.74}$ $\times$ 10$^{-12}$ & 2.33$^{+0.39}_{-0.79}$ $\times$ 10$^{-12}$ & 2.28$^{+0.45}_{-0.81}$ $\times$ 10$^{-12}$ & 2.24$^{+0.15}_{-0.93}$ $\times$ 10$^{-12}$ & 2.17$^{+0.53}_{-2.63}$ $\times$ 10$^{-12}$ \\
L$\rm_{2-10\,keV}$ & 7.75$^{+1.72}_{-3.49}$ $\times$ 10$^{41}$ & 7.55$^{+9.05}_{-2.00}$ $\times$ 10$^{41}$ & 8.28$^{+1.25}_{-0.69}$ $\times$ 10$^{41}$ & 8.32$^{+1.01}_{-0.91}$ $\times$ 10$^{41}$ & 9.71$^{+92.29}_{-2.93}$ $\times$ 10$^{41}$\\
L$\rm_{15-55\,keV}$ & 2.31$^{+0.51}_{-1.04}$ $\times$ 10$^{42}$ & 2.67$^{+3.21}_{-0.71}$ $\times$ 10$^{42}$ & 1.00$^{+0.10}_{-0.88}$ $\times$ 10$^{42}$ & 1.86$^{+0.14}_{-0.17}$ $\times$ 10$^{42}$ & 1.96$^{+18.54}_{-0.59}$ $\times$ 10$^{42}$ \\
\enddata
\vspace{5mm}
\textbf{Notes:} Same as Table \ref{tab:2m02nust}. Additional parameters: \\ 
kT: \texttt{mekal} model temperature in units of keV. \\
c$_{f,Tor}$: Covering factor of the torus. \\
CTKcover: Covering factor of the inner ring of clouds, computed with \texttt{UXClumpy}. \\
Tor $\sigma$: Cloud dispersion factor, computed with \texttt{UXClumpy}. \\
cos($\theta_{obs}$): Cosine of the inclination angle. \\
$N_{H, avg}$: Average torus hydrogen column density, in units of 10$^{24}$ cm$^{−2}$. \\
$f_s$: Fraction of scattered continuum. \\
*: Indicates the parameter was frozen to this value during fitting. \\
u: The parameter is unconstrained.
\end{deluxetable*}

\begin{deluxetable*}{ccccccc} [h!]

\tablecaption{ESO 362$-$8 \nus -\xmm}
\label{tab:eso362}

\tablehead{\colhead{Model} & \colhead{MYTorus} & \colhead{MYTorus} & \colhead{borus02} & \colhead{borus02} & \colhead{UXClumpy} \\ 
\colhead{} & \colhead{(Decoupled Face-on)} & \colhead{(Decoupled Edge-on)} & & \colhead{Two $\Gamma$ }}

\startdata
$\chi^2$/dof & 71/82 & 72/82 & 80/81 & 71/80 & 68/80 \\
kT & 0.66$^{+0.10}_{-0.08}$ & 0.66$^{+0.11}_{-0.09}$ & 0.62$^{+0.06}_{-0.07}$ & 0.65$^{+0.07}_{-0.08}$ & 0.77$^{+0.08}_{-0.08}$ \\
$\Gamma$ & 1.90$^{+0.17}_{-0.22}$ & 1.41$^{+0.08}_{-u}$ & 2.6$^{+u}_{-0.31}$ & 1.80$^{+0.44}_{-0.22}$ & 1.68$^{+0.13}_{-0.15}$\\
norm $10^{-2}$ & 0.14$^{+0.03}_{-0.05}$ & 0.05$^{+0.01}_{-0.02}$ & 2.14$^{+0.60}_{-0.10}$ & 0.09$^{+1.81}_{-0.08}$ & 0.29$^{+1.54}_{-0.19}$\\
c$_{f,Tor}$ & ... & ... & 0.90$^{+0.08}_{-0.02}$ & 0.91$^{+0.08}_{-0.16}$ & ... \\
CTKcover & ... & ... & ... & ... & 0.31$^{+u}_{-0.11}$ \\
Tor $\sigma$ & ... & ... & ... & ... & 14.0$^{+u}_{-4.7}$ \\
cos($\theta_{obs}$) & ... & ... & 0.85$^{+0.04}_{-0.08}$ & 0.89$^{+u}_{-0.09}$ & 1.00$^{+u}_{-u}$\\
$N_{H, l.o.s.}$ & 2.78$^{+u}_{-0.65}$ & 2.18$^{+0.21}_{-0.30}$ & 3.96$^{+u}_{-1.30}$ & 2.57$^{+1.40}_{-0.80}$ & 3.93$^{+u}_{-1.41}$ \\
$N_{H, avg}$ & 9.91$^{+u}_{-5.66}$ & 0.15$^{+0.11}_{-0.04}$ & 10$^{+0.15}_{-4.75}$ & 31.62$^{+u}_{-11.11}$ & ... \\
$f_s$ 10$^{-2}$ & 0.90$^{+0.80}_{-0.30}$ & 2.60$^{+1.00}_{-0.60}$ & 0.05$^{+0.01}_{-0.02}$ & 1.40$^{+3.60}_{-1.30}$ & 1.20$^{+5.50}_{-1.30}$ \\
$\Gamma$ \#2 & 3.03$^{+0.35}_{-0.41}$ & 2.94$^{+0.32}_{-0.37}$ & ... & 2.92$^{+0.38}_{-0.55}$ & 3.00$^{+u}_{-0.26}$ \\ 
c$_{nus}$ & 1.07$^{+0.29}_{-0.20}$ & 1.12$^{+0.30}_{-0.19}$ & 1.14$^{+0.30}_{-0.25}$ & 1.10$^{+0.25}_{-0.23}$ & 1.07$^{+0.26}_{-0.21}$ \\
\hline
F$\rm_{2-10\,keV}$ & 9.79$^{+0.77}_{-2.54}$ $\times$ 10$^{-14}$ & 1.04$^{+3.87}_{-4.23}$ $\times$ 10$^{-13}$ & 9.39$^{+1.67}_{-5.02}$ $\times$ 10$^{-14}$ & 1.00$^{+3.12}_{-5.63}$ $\times$ 10$^{-13}$ & 1.02$^{+2.46}_{-5.32}$ $\times$ 10$^{-13}$ \\
F$\rm_{15-55\,keV}$ & 2.19$^{+1.01}_{-0.71}$ $\times$ 10$^{-12}$ & 2.04$^{+1.81}_{-0.99}$ $\times$ 10$^{-12}$ & 1.73$^{+0.78}_{-1.23}$ $\times$ 10$^{-12}$ & 2.17$^{+3.87}_{-3.83}$ $\times$ 10$^{-12}$ & 2.19$^{+3.45}_{-4.23}$ $\times$ 10$^{-12}$ \\
L$\rm_{2-10\,keV}$ & 2.06$^{+0.41}_{-0.75}$ $\times$ 10$^{42}$ & 1.71$^{+0.22}_{-0.74}$ $\times$ 10$^{42}$ & 4.07$^{+11.78}_{-u}$ $\times$ 10$^{43}$ & 2.82$^{14.96}_{-u}$ $\times$ 10$^{43}$ & 5.43$^{+31.67}_{-3.54}$ $\times$ 10$^{42}$\\
L$\rm_{15-55\,keV}$ & 2.34$^{+0.47}_{-0.85}$ $\times$ 10$^{42}$ & 4.91$^{+0.65}_{-2.13}$ $\times$ 10$^{42}$ & 2.51$^{+2.06}_{-2.01}$ $\times$ 10$^{42}$ & 1.20$^{+0.35}_{-0.37}$ $\times$ 10$^{43}$ & 1.05$^{+5.64}_{-0.69}$ $\times$ 10$^{43}$ \\
\enddata
\vspace{5mm}
\textbf{Notes:} Same as Table \ref{tab:2m04nus}. \\ 
\end{deluxetable*}

\begin{deluxetable*}{cccccc} [h!]

\tablecaption{IC 2227 \nus - \xmm}
\label{tab:ic2227}

\tablehead{\colhead{Model} & \colhead{MYTorus} & \colhead{MYTorus} & \colhead{borus02} & \colhead{UXClumpy} \\ 
\colhead{} & \colhead{(Decoupled Face-on)} & \colhead{(Decoupled Edge-on)} }

\startdata
$\chi^2$/dof & 165/140 & 180/140 & 165/138 & 172/138 \\
kT & 0.63$^{+0.06}_{-0.07}$ & 0.62$^{+0.06}_{-0.07}$ & 0.63$^{+0.06}_{-0.06}$ & 0.75$^{+0.07}_{-0.09}$ \\
$\Gamma$ & 1.86$^{+0.20}_{-0.14}$ & 1.75$^{+0.23}_{-0.22}$ & 1.87$^{+0.12}_{-0.11}$ & 1.95$^{+0.19}_{-0.04}$ \\
norm $10^{-2}$ & 0.09$^{+0.09}_{-0.03}$ & 0.07$^{+0.09}_{-0.04}$ & 0.09$^{+0.05}_{-0.03}$ & 0.13$^{+0.01}_{-0.06}$ \\
c$_{f,Tor}$ & ... & ... & 0.80$^{+0.07}_{-0.11}$ & ... \\
CTKcover & ... & ... & ... & 0.60$^{+u}_{-0.11}$ \\
Tor $\sigma$ & ... & ... & ... & 16.1$^{+57.2}_{-12.6}$  \\
cos($\theta_{obs}$) & ... & ... &  0.78$^{+0.12}_{-0.02}$ & 1.00$^{+u}_{-0.59}$ \\
$N_{H, l.o.s.}$ & 0.60$^{+0.11}_{-0.07}$ & 0.58$^{+0.08}_{-0.08}$ &  0.64$^{+0.05}_{-0.06}$ & 0.63$^{+0.10}_{-0.04}$ \\
$N_{H, avg}$ & 6.06$^{+u}_{-2.84}$ & 1.37$^{+1.96}_{-1.24}$ &  31.62$^{+u}_{-26.00}$ & ... \\
$f_s$ 10$^{-2}$ & 0.90$^{+0.50}_{-0.40}$ & 1.10$^{+1.10}_{-0.50}$ & 0.90$^{+0.60}_{-0.10}$ & 3.20$^{+3.00}_{-1.00}$ \\ 
c$_{nus}$ & 1.03$^{+0.11}_{-0.10}$ & 0.99$^{+0.11}_{-0.10}$ & 1.03$^{+0.12}_{-0.06}$ & 1.02$^{+0.11}_{-0.11}$ \\
\hline
F$\rm_{2-10\,keV}$ & 3.36$^{+0.51}_{-0.87}$ $\times$ 10$^{-13}$ & 3.45$^{+0.39}_{-0.72}$ $\times$ 10$^{-13}$ & 3.37$^{+0.43}_{-1.09}$ $\times$ 10$^{-13}$ & 3.40$^{+0.27}_{-1.64}$ $\times$ 10$^{-13}$\\
F$\rm_{15-55\,keV}$ & 3.00$^{+0.98}_{-1.60}$ $\times$ 10$^{-12}$ & 2.72$^{+0.52}_{-0.77}$ $\times$ 10$^{-12}$ & 2.93$^{+0.23}_{-1.32}$ $\times$ 10$^{-12}$ & 2.74$^{+0.88}_{-1.46}$ $\times$ 10$^{-12}$ \\
L$\rm_{2-10\,keV}$ & 5.60$^{+6.10}_{-1.70}$ $\times$ 10$^{42}$ & 5.74$^{+6.76}_{-3.40}$ $\times$ 10$^{42}$ & 6.31$^{+2.81}_{-0.94}$ $\times$ 10$^{42}$ & 6.81$^{+0.66}_{-3.07}$ $\times$ 10$^{42}$ \\
L$\rm_{15-55\,keV}$ & 6.58$^{+7.22}_{-1.99}$ $\times$ 10$^{42}$ & 7.92$^{+9.28}_{-4.69}$ $\times$ 10$^{42}$ & 2.95$^{+1.62}_{-0.95}$ $\times$ 10$^{42}$ & 8.77$^{+0.67}_{-4.06}$ $\times$ 10$^{42}$ \\
\enddata
\vspace{5mm}
\textbf{Notes:} Same as Table \ref{tab:2m04nus}. \\ 
\end{deluxetable*}

\bibliography{bibliography}

\end{document}